%Paper: hep-th/9410211
%From: Jacquot Jean-Luc <jacquot@crnhp4.in2p3.fr>
%Date: Thu, 27 Oct 94 18:23:01 MET

%%%%% definitions &&&&&&&&&&&&&&&&&&&&&
 %\input texccs>listing
%\listing{routine}
\font\smc=cmcsc10 scaled 913

%\font\lins=line10
\font\cmis=cmsltt10 scaled 833

%\font\fdivi=cmssmc40 scaled 833
\font\slz=cmex10 scaled 1440
\font\sqi=cmssq8

\def\lrto{\Longleftrightarrow}
%
%   mathleg
%

%

%
% scriptstyle et displaystyle
%
\def\sty{\scriptstyle}

\def\dty{\displaystyle}
%
% a commercial

%
%  fleche de longueur variable
%
\def\fle{\hbox to 35pt{\rightarrowfill}}
%
%  flechesh plus courte
%

%

%

%/ zero      \znot

%
\def\cun{$\bigcirc\kern-7.pt {\scriptstyle 1}\hskip 7pt$}
\def\cdeux{$\bigcirc\kern-7.pt {\scriptstyle 2}\hskip 7pt$}
\def\ctrois{$\bigcirc\kern-7.pt {\scriptstyle 3}\hskip 7pt$}
\def\cquatre{$\bigcirc\kern-7.pt {\scriptstyle 4}\hskip 7pt$}
\def\ccinq{$\bigcirc\kern-7.pt {\scriptstyle 5}\hskip 7pt$}
\def\csix{$\bigcirc\kern-7.pt {\scriptstyle 6}\hskip 7pt$}
\def\csept{$\bigcirc\kern-7.pt {\scriptstyle 7}\hskip 7pt$}
\def\chuit{$\bigcirc\kern-7.pt {\scriptstyle 8}\hskip 7pt$}
\def\cneuf{$\bigcirc\kern-7.pt {\scriptstyle 9}\hskip 7pt$}
\def\czero{$\bigcirc\kern-7.pt {\scriptstyle 0}\hskip 7pt$}

\def\caa{$\bigcirc\kern-7.pt {\scriptstyle \rm a}\hskip 7pt$}
\def\cbb{$\bigcirc\kern-7.pt {\scriptstyle \rm b}\hskip 7pt$}
\def\ccc{$\bigcirc\kern-7.pt {\scriptstyle \rm c}\hskip 7pt$}
%
% tau plus grand
% Psi plus grand
% slash pour DA
% sim plus grand
% ›ell + grand et ›pi + grand
%
%\def\wtil{{\sly\char '030}}
\def\wtil{{\slz\char '146}}
\def\bsla{{\cmis\char '134}}

\newbox\prov
\def\ssa#1#2{\global\setbox\prov=\hbox{$\underbrace{#1}_{#2}$}}
\def\tilda#1{\setbox0=\hbox{{\tenex\char'0145}}%
       \wd0=0pt \lower0.5\ht0\box0#1}
%
%overrightarrow plus court
%  overline vep gui dif
%

\def\gui{\char'134 }

%
% accolades align{es
%
\def\accolad#1#2{$$\hbox to \hsize{\rlap{\hbox{#2}}
                                   \hskip 1truein $#1$
                                   \hfill}$$}
\def\bccolad#1#2{$$\hbox to \hsize{\rlap{\hbox{#2}}
                                   \hskip 1truecm $#1$
                                   \hfill}$$}
%
%  N+VERT
%
\def\DR{\rm I\kern-1.45pt\rm R}
\def\DN{\rm I\kern-1.45pt\rm N}
\def\DZ{\hbox{$\rm Z$}
        \hbox{\kern -4pt \rm Z}}
\def\DZP{\hbox{$\sty Z$}
        \hbox{\kern -4pt $\sty Z$}}
\def\DF{\rm 1\kern-2.95pt\rm I}
\def\DU{\rm 1\kern-2.95pt\rm I}
\def\DC{{\sqi I}\kern-4.2pt\rm C}
\def\DA{{\rm A}\kern-3.9pt\raise 1pt \hbox{\bsla}}
\def\DAA{\rm A \kern-10pt\rm A}
\def\DI{\rm 1\kern-2.9pt\rm I}
%
%subset +fleche
\def\subto{\hbox{$\subset$}
     \hbox{\kern -2.5pt${}_\gt$\ }}
%
%subset+2int+supset
\def\bigoint{\hbox{$\;\;\subset$
      \kern -12.0pt$\dty\int$
      \kern -9.0pt$\dty\int$
      \kern -12.0pt$\supset$\ }}
%subset+3int+supset
\def\biggoint{\hbox{$\;\;\;\subset$
      \kern -13.0pt$\dty\int$
      \kern -9.45pt$\dty\int$
      \kern -9.45pt$\dty\int$
      \kern -12.5pt$\supset$\ }}
%
% carre noir

%
%trinoir
%
\newdimen\dtrinoir\dtrinoir=0pt
\def\trinoir{\dtrinoir=0 pt
    \loop\ifdim\dtrinoir<10pt
         \advance\dtrinoir by 0.25pt
         \vrule height\dtrinoir width .125pt depth 0pt
         \hskip 0pt
         \repeat
    \loop\ifdim\dtrinoir>0pt
         \advance\dtrinoir by -.25pt
         \vrule height\dtrinoir width .125pt depth 0pt
         \hskip 0pt
         \repeat}
%
% majuscules calligraphe
%

\def\da{{\cal D}}

%
%    trait de 75pt
%
\def\fla{\hbox to 75pt{\rightarrowfill}}
%
%    vide + petit que hauteur
%

%
%
%    \remplitrait
%    \remplivide
%

%
%  def tita
%
%\def\tita#1{\goodbreak\vskip 1truecm \noindent
%             {\bf #1}\nobreak\vskip .8truecm}
%
%  def titb
%
%\def\titb#1{\goodbreak\bigskip\noindent
%             {\slbf #1}\nobreak\medskip}
%
%  def titc
%
%\def\titc#1{\goodbreak\medskip\noindent
%             {\it #1}\nobreak\smallskip}
%
%  def titt
%
%\def\titt#1{\goodbreak\bigskip\noindent
%             {\underbar{\bf#1}}\nobreak\medskip}
%
%   footnote
%
\newcount\numnote
\def\note#1.#2*{\ifnum#1>\numnote \numnote=#1\footnote{$|#1$}{#2}
                  \else$|#1$
                  \fi}
%
% fleche brisee
%
\def\fleche#1#2{\setbox2=\hbox{$#1$}
                \mathop#1_{\hskip .3\wd2
               \raise .5ex \hbox{\vrule height 12pt}
%                \hbox{\vrule height 12pt}
                \rlap{$\mkern-1.5mu\rightarrow#2$}}}
%\def\fleche#1#2{\setbox2=\hbox{$#1$}
%                \mathop#1\limits_{\hskip .3\wd2
%                \raise .5ex \hbox{\vrule height 12pt}
%                \rlap{$\mkern-1.5mu\rightarrow{\rm #2}$}}}
%
% fleche brisee vers le haut
%

\def\flecha#1#2{\setbox2=\hbox{$#1$}
                \mathop{#1}_{\hskip .3\wd2
               \raise .5ex \hbox{\vrule height 24pt}
                \rlap{$\mkern-1.5mu\rightarrow#2$}}}
%
% fleche brisee  sans premier argument
%

%
% longue fleche leftrightarrow
%
\def\lrfle{\leftarrow\kern-2pt\rightarrow}
\def\Lrfle{\longleftarrow\kern-3pt\longrightarrow}
\def\ddds{\mathinner{\mskip1mu\raise4pt\vbox{\kern4pt\hbox{.}}\mskip1mu
\raise4pt\hbox{.}\mskip1mu\raise1pt\hbox{.}\mskip1mu}}
\newbox\crov
\newdimen\crovd
\setbox\crov=\hbox{$(1+\bar c\bar z)$}
\crovd=1\wd\crov
\advance \crovd by 4pt
\wd\crov=\crovd
\def\lin{\hbox to \crovd{\hrulefill}}

\newbox\brov
\newdimen\brovd
\setbox\brov=\hbox{$(1+cz)$}
\brovd=1\wd\brov
\advance \brovd by 4pt
\wd\brov=\brovd
\def\lin{\hbox to \brovd{\hrulefill}}

%
% petites matrices sans blanc
%
\def\matmu{
          \hbox{\vbox{\hbox{$\sty\mu$}
                      \kern-4.5pt
                      \hbox{$\cdot$}}
                \vbox{\hbox{$.$}
                      \kern-4.5pt
                      \hbox{$\sty\mu$}}}
          }
\def\matmunu{
          \hbox{\vbox{\hbox{$.$}
                      \kern-4.5pt
                      \hbox{$\sty\mu$}}
                \vbox{\hbox{$\sty\nu$}
                      \kern-4.5pt
                      \hbox{$\cdot$}}}
          }
%
%
%

%
% fleches croisees
%

%
% longue fleche verticale

%
%     lt et gt
%
\def\lt{\lt}
\def\gt{\gt}
%
%  crochet ouvrant           \cg
%  crochet fermant           \cd
%
\def\cg{\lbrack}
\def\cd{\rbrack}
\def\cg{\char'133 }
\def\cd{\char'135 }
%
% ecrire un area sur deux colonnes
%
\newbox\gauche \newbox\droite
\long\def\zona #1#2{
               \setbox\gauche=\vtop{\hsize=3.2truein #1\par}
               \setbox\droite=\vtop{\hsize=3.4truein #2\par}
               \line{\box\gauche\hfill\box\droite}}
\newbox\gauche \newbox\droite
\long\def\zonb #1#2{
               \setbox\gauche=\vtop{\hsize=1.5truein #1\par}
               \setbox\droite=\vtop{\hsize=3.7truein #2\par}
               \line{\box\gauche\hfill\vrule\hfill\box\droite}}
\newbox\gauche \newbox\droite
\long\def\zonc #1#2{
               \setbox\gauche=\vtop{\hsize=1.90truein #1\par}
               \setbox\droite=\vtop{\hsize=4.00truein #2\par}
               \line{\box\gauche\hfill\box\droite}}
%
% ecrire un area sur deux colonnes
%
\newbox\gauche \newbox\droite
\long\def\zone #1#2{
               \setbox\gauche=\vtop{\hsize=2.75truein #1\par}
               \setbox\droite=\vtop{\hsize=2.85truein #2\par}
               \line{\box\gauche\hfill\box\droite}}
\newbox\gauche \newbox\droite
\long\def\zontdt#1#2{
               \setbox\gauche=\vtop{\hsize=2.75truein #1\par}
               \setbox\droite=\vtop{\hsize=2.85truein #2\par}
               \line{\box\gauche\hfill\vrule\hfill\box\droite}}
\newbox\gauche \newbox\droite
\long\def\zonf #1#2{
               \setbox\gauche=\vtop{\hsize=3.5truein #1\par}
               \setbox\droite=\vtop{\hsize=2truein #2\par}
               \line{\box\gauche\hfill\box\droite}}
\newbox\gauche \newbox\droite
\long\def\zonk #1#2{
               \setbox\gauche=\vtop{\hsize=2.5truein #1\par}
               \setbox\droite=\vtop{\hsize=2.9truein #2\par}
               \line{\box\gauche\hfill\box\droite\hfill}}
%
% ecrire un area sur deux colonnes
%
\newbox\gauche \newbox\droite
\long\def\zonw #1#2{
               \setbox\gauche=\vtop{\hsize=2.75truein #1\par}
               \setbox\droite=\vtop{\hsize=2.75truein #2\par}
               \line{\box\gauche\hfill\box\droite}}
%
% ecrire un area sur deux colonnes
%
\newbox\gauche \newbox\droite
\long\def\zont #1#2{
               \setbox\gauche=\vtop{\hsize=0.3truein #1\par}
               \setbox\droite=\vtop{\hsize=5.6truein #2\par}
               \line{\box\gauche\hfill\box\droite}}
\newbox\gauche \newbox\droite
\newbox\milieu
\long\def\zontt#1#2#3{
               \setbox\gauche=\vtop{\hsize=0.3truein #1\par}
               \setbox\milieu=\vtop{\hsize=2.5truein #2\par}
               \setbox\droite=\vtop{\hsize=2.5truein #3\par}
               \line{\box\gauche\hfill\box\milieu\hfill\box\droite}}
%
%lambda bar=lambar
%

%
%
%
%  o trema                   \otrema

%
%  a trema                   \atrema

%
% sim dans J
%   -  dans J                \J

%\def\J{\csim\mskip-3.6mu\char'112}
%
%      superieur et equivalent
\def\supsim{\mathaccent"013E{\lower1ex\hbox{$\mathchar"3218$}}}
%
%      inferieur et equivalent
\def\infsim{\mathaccent"013C{\lower1ex\hbox{$\mathchar"3218$}}}
%
%  Dallenbertien             \dal

%
\mathchardef\lt="313C
\mathchardef\gt="313E
%
%      > et tilda
%

%
%      lettre sur tilda
%
\def\surtilda#1{\setbox0=\hbox{#1}\setbox2=\hbox{\~{}}
    \ifdim\wd0>\wd2
     \vtop{\baselineskip 1pt\hbox{#1}\hbox to1\wd0{\hfil\~{}\hfil}}
    \else
     \vtop{\baselineskip 1pt\hbox to\wd2{\hfil#1\hfil}\hbox{\~{}}}
    \fi}
%
%      lettre sur tilda
%
\def\surtildo#1{\setbox0=\hbox{#1}\setbox2=\hbox{$\eightpoint\sim${}}
    \ifdim\wd0>\wd2
     \vtop{\baselineskip 1pt\hbox{#1}\hbox
    to1\wd0{\hfil$\eightpoint\sim${}\hfil}}
    \else
     \vtop{\baselineskip 1pt\hbox
    to\wd2{\hfil#1\hfil}\hbox{$\eightpoint\sim${}}}
    \fi}
%
%      lettre sur grand tilda
%  demande wtil et font slz (ou sly)
%
\def\surtildg#1{\setbox0=\hbox{#1}\setbox2=\hbox{$\wtil${}}
    \ifdim\wd0>\wd2
     \vtop{\baselineskip 3pt\hbox{#1}\hbox
    to1\wd0{\hfil\wtil{}\hfil}}
    \else
     \vtop{\baselineskip 3pt\hbox
    to\wd2{\hfil#1\hfil}\hbox{\wtil{}}}
    \fi}
%
%      lettre sous fleche
%
\def\soufle{\setbox2=\hbox{$\eightpoint\longleftrightarrow$}
     \vbox{\baselineskip 1pt
      \hbox{$\eightpoint\longleftrightarrow$}
      \hbox to\wd2{\hfill$\partial_\mu$\hfill}}}
%

%
%     brack sur lettre
%
\def\hb{\vbox{\hrule\hbox{\vrule depth 4pt
            \hbox to .85truecm{\ \hfill}\vrule}}}

\def\hc{\vbox{\hrule\hbox{\vrule depth 4pt
            \hbox to .60truecm{\ \hfill}\vrule}}}

%
%     entete de page
%
%\headline={\tenrm\hfil\entete\hfil}
%\def\entete{}
%\def\makeheadline{\vbox to 0pt{\vskip-42.5pt
%    \hbox to 6.5truein{\vbox to8.5pt{}\the\headline}\vss}
%                       \nointerlineskip}
%
% \un5.Niveau 1\par
% \deux5.2.Niveau 5 2\par
% \deu5.2.Niveau 5 2\par
% \trois5.2.4.Niveau 5 2 4\par
% \quatre5.2.4.6.Niveau 5 2 4 6\par
%
\outer\def\un#1.#2\par{%
                       {\bf
                        \noindent \null \hskip -1 em
                        #1\enspace#2 \par}
                       \nobreak\medskip}

\outer\def\deux#1.#2.#3\par{%
                 {\ninepoint\bf
                 \textfont0=\tenbf
                 \textfont1=\tenbi
                 \textfont2=\tenbsy
                             \noindent \null \hskip -1 em
                             #1.#2\enspace#3\par\tenpoint}
                            \nobreak\medskip}

\outer\def\trois#1.#2.#3.#4\par{%
                \numnote=0
                {\bigskip\bf
                 \noindent \null \hskip -1 em
                 #1.#2.#3\enspace#4 \par\tenpoint}
                \nobreak \medskip}
\outer\def\quatre#1.#2.#3.#4.#5\par{%
                {\it
                 \noindent \null \hskip -1 em
                 #1.#2.#3.#4\enspace#5 \par}
                \nobreak \medskip}
%\outer\def\quatre#1.#2.#3.#4.#5\par{%
%               \numnote=0
%                {\bf
%                 \textfont0=\tenbf
%                 \textfont1=\tenbi
%                 \textfont2=\tenbsy
%                 \noindent \null \hskip -1 em
%                 #1.#2.#3.#4\enspace#5 \par}
%                \nobreak \medskip}
%
%  \abstract  resume   \endabstract
%
%\newskip\margabstract
%\margabstract=2truecm
%\def\abstract{\par
%              \bigskip
%              \begingroup
%              \it
%              \leftskip=\margabstract
%              \rightskip=\margabstract
%              \ignorespaces}
%\def\endabstract{\par
%                 \endgroup
%                 \bigskip}
%
%  \obstract  resume   \endabstract
%
%\newskip\margabstract
%\margabstract=2truecm
%\def\obstract{\par
%              \medskip
%              \begingroup
%              \leftskip=\margabstract
%              \rightskip=\margabstract
%              \ignorespaces}
%\def\endabstract{\par
%                 \endgroup
%                 \bigskip}
\newskip\margabstract
\margabstract=2truecm
\newskip\margobstract
\margobstract=1.5truecm
\def\abstract{\par
              \anglais
              \bigskip
              \begingroup
              \it
              \leftskip=\margabstract
              \rightskip=\margabstract
              \ignorespaces}

%
%  \collaboration  liste\par
%
\def\collaboration#1\par{\medskip
                         \noindent
                         COLLABORATION
                         \par
                         \nobreak
                         \smallskip
                         \noindent#1
                         \par
                         \smallskip}
%
%  \Collaboration<nom>  liste\par
%
\def\Collaboration#1#2\par{\medskip
                         \noindent
                         COLLABORATION #1
                         \par
                         \nobreak
                         \smallskip
                         \noindent#2
                         \par
                         \smallskip}
%
%  \noms liste\par
%
\def\noms#1\par{\medskip
               \smallskip
               \noindent{#1}
               \par
               \smallskip}
%
%  \ref(5)
%
\def\ref(#1){{\lbrack}#1{\rbrack}}
%
%  \note5.ULP*
%            ou \note2.*
%
\newcount\numnote
\def\note#1.#2*{\ifnum#1>\numnote \numnote=#1\footnote{$|#1$}{#2}
                  \else$|#1$
                  \fi}
%
%  \Ref
% ou
%  \Refs
%
%      \aut nom*
%      \auteur nom*
%      \livre ....*
%      \editeur ...*
%      \gras  ....*
%
\newcount\numpub

\def\aut#1*{\advance \numpub by 1
            \smallskip
            \hangindent 1.5 truecm
            \noindent \hskip 1 truecm
            \llap{\cg\the\numpub\cd\hskip.35em}
            {\smc#1}}
\def\autu#1*{\advance \numpub by 1
            \smallskip
            \hangindent 1.5 truecm
            \noindent \hskip 1 truecm
            \llap{\cg1.\the\numpub\cd\hskip.35em}
            {\smc#1}}
\def\autd#1*{\advance \numpub by 1
            \smallskip
            \hangindent 1.5 truecm
            \noindent \hskip 1 truecm
            \llap{\cg2.\the\numpub\cd\hskip.35em}
            {\smc#1}}
\def\autt#1*{\advance \numpub by 1
            \smallskip
            \hangindent 1.5 truecm
            \noindent \hskip 1 truecm
            \llap{\cg3.\the\numpub\cd\hskip.35em}
            {\smc#1}}
\def\suitaut#1*{\par
                \hangindent 1.5 truecm
                \noindent \hskip 1 truecm
                {\smc#1}}
\def\auteur#1*{\smallskip
            \hangindent 1.5 truecm
            \noindent
            {\smc#1}}
\def\livre#1*{\enskip ---\enskip #1\ignorespaces}
\def\editeur#1*{\enskip ---\enskip {\it #1}}
\def\gras#1*{\ {\bf #1} } %
%
%    \math(5)
%
%    \Math(5)
%
\def\math(#1){$\bf\backslash math(#1)$\ \wlog{math (#1)}}
\def\Math(#1){$$\bf\backslash Math(#1)$$\ \wlog{Math (#1)}}
%
%    \image(largeur)(hauteur)<legende>
%
\def\image(#1)(#2)#3{\goodbreak
                      \midinsert
                         \vskip #2 truecm
                         \hsize #1 truecm
                         #3
                       \endinsert}
\def\nv{\nobreak\kern -4pt,\ }
%
% petits points
%
\def\leaderfill{\leaders\hbox to 1em{\hss.\hss}\hfill}
\mathchardef\lt="313C
\mathchardef\gt="313E
%
%  Dallenbertien             \Dal  \Dam

%  /partial                  \spa
\def\spa{\setbox0=\hbox{$\partial$}\partial\hskip-1.65\wd0\not}
%
%slah sur lettre
% D P A a b q p k
%
\def\Dbar{{\char'057\mskip-13mu\char'104}}

\def\pbar{{\char'057\mskip-10mu\char'160}}

%
% sinus cosinus et tangente hyperb.
%

%
%     noindent raccourci
%
\def\noi{\noindent}
%
% prime raccourci et seconde ›pr et ›prr

%
%
\def\nto{{}_{{}_{\to}}\kern-7.5pt n \kern 5pt}
\def\nta{{{}\atop \to} \kern-9pt n \kern 4pt }
\def\nti{{{}\atop{ \to\atop{}}} \kern-9pt n \kern 4pt}
% skip 10pt$= \cg\vep\cd\;(\quad \cdot\quad;
% ovi u_1;\cdots \ovi u_{n-1})
% }
%
% petites matrices sans blanc
%
\def\mat#1#2{
          \hbox{\vbox{\hbox{$\sty#1$}
                      \kern-8.5pt
                      \hbox{$\cdot$}}
                \vbox{\hbox{$.$}
                      \kern-8.5pt
                      \hbox{$\sty#2$}}}
          }
\def\mmat#1#2{
          \hbox{\vbox{\hbox{$\sty#1$}
                      \kern-8.5pt
                      \hbox{$~$}}
                \vbox{\hbox{$~$}
                      \kern-8.5pt
                      \hbox{$\sty#2$}}}
          }
\def\tmat#1#2#3{
          \hbox{\vbox{\hbox{$\sty#1$}
                      \kern-8.5pt
                      \hbox{$\cdot$}}
                \vbox{\hbox{$\sty#2$}
                      \kern-8.5pt
                      \hbox{$\cdot$}}
                \vbox{\hbox{$.$}
                      \kern-8.5pt
                      \hbox{$\sty#3$}}}
          }
\def\pat#1#2{
          \hbox{\vbox{\hbox{$\cdot$}
                      \kern-8.5pt
                      \hbox{$\sty#2$}}
                \vbox{\hbox{$\sty#1$}
                      \kern-8.5pt
                      \hbox{$.$}}}
          }
\def\tpat#1#2#3{
          \hbox{\vbox{\hbox{$\cdot$}
                      \kern-8.5pt
                      \hbox{$\sty#3$}}
                \vbox{\hbox{$\cdot$}
                      \kern-8.5pt
                      \hbox{$\sty#2$}}
                \vbox{\hbox{$\sty#1$}
                      \kern-8.5pt
                      \hbox{$.$}}}
          }
\def\upat#1#2#3#4{
          \hbox{\vbox{\hbox{$\sty#4$}
                      \kern-8.5pt
                      \hbox{$.$}}
                \vbox{\hbox{$\cdot$}
                      \kern-8.5pt
                      \hbox{$\sty#3$}}
                \vbox{\hbox{$\cdot$}
                      \kern-8.5pt
                      \hbox{$\sty#2$}}
                \vbox{\hbox{$\cdot$}
                      \kern-8.5pt
                      \hbox{$\sty#1$}}}
          }
\def\mata{
          \hbox{\vbox{\hbox{$\sty\mu$}
                      \kern-8.5pt
                      \hbox{$\cdot$}}
                \vbox{\hbox{$.$}
                      \kern-8.5pt
                      \hbox{$\sty\nu$}}}
          }
\def\matb{
          \hbox{\vbox{\hbox{$\sty0$}
                      \kern-8.5pt
                      \hbox{$\cdot$}}
                \vbox{\hbox{$.$}
                      \kern-8.5pt
                      \hbox{$\sty i$}}}
          }
\def\matc{
          \hbox{\vbox{\hbox{$\sty\mu$}
                      \kern-8.5pt
                      \hbox{$\cdot$}}
                \vbox{\hbox{$.$}
                      \kern-8.5pt
                      \hbox{$\sty 0$}}}
          }
%
%      superieur et equivalent
\def\supsim{\mathaccent"013E{\lower1ex\hbox{$\mathchar"3218$}}}
%
%      inferieur et equivalent
\def\infsim{\mathaccent"013C{\lower1ex\hbox{$\mathchar"3218$}}}
%

%
% begining of the paper %%%%%%%%%%%%%%%%%%%%%%%%%%%%%
\magnification=\magstep1
\baselineskip=14pt
\hsize=6truein
\vsize=9truein
\parindent=.65truecm
\def\tita#1{\goodbreak\vskip .9truecm\noindent
            #1\nobreak\bigskip}
\def\gui{\char'134}
\def\da{{\cal D}}
\def\Dbar{{\char'057\mskip-13mu\char'104}}
\def\spa{\setbox0=\hbox{$\partial$}\partial\hskip-1.65\wd0\not}
\def\editeur#1*{:\enskip {\it #1}}

\null\vskip1truecm

\centerline{\bf Equations from Non-Linear Chiral}
\centerline{\bf Transformations}
\vskip1truecm
\centerline{J.L. Jacquot, J. Richert and M. Umezawa}
\vskip.5truecm
\centerline{\it Centre de Recherches Nucl\'eaires}
\centerline{\it IN2P3-CNRS / Universit\'e Louis Pasteur}
\centerline{\it BP 20, F-67037 STRASBOURG Cedex 2, FRANCE}
\vskip2.5truecm

Abstract
\vskip1truecm

\vbox{\baselineskip=20pt
In comparison with the $WT$ chiral identity which is indispensable for
renormalization theory, relations deduced from the non-linear chiral
transformation have a totally different physical significance.
We wish to show that non-linear chiral transformations are powerful tools
to deduce useful integral equations for propagators. In contrast to the
case of linear chiral transformations, identities derived from non-linear ones
contain
more involved radiative effects and are rich in physical content.
To demonstrate this fact we apply the simplest non-linear chiral
transformation to the Nambu-Jona-Lasinio model, and show how our identity
is related to the Dyson-Schwinger equation and Bethe-Salpeter amplitudes
of the Higgs and $\pi$. Unlike equations obtained from the effective
potential, our resultant equation is exact and can be used for events
beyond the LEP energy}

\vfill\eject
\tita{1.  Introduction}

Over the past four years Tevatron and LEP experiments have stimulated
investigations
on models beyond the standard model. In addition to those in existence for many
years (preon,
technicolor, supersymmetry, etc.) a new idea has been proposed to explain the
heaviness of the top mass by a still unknown Nambu-Jona-Lasinio interaction
(NJL) [1], which itself is an effective interaction of also a yet unknown
vector meson based upon the underlying philosophy named \gui Bootstrap"
by Nambu [2]. When such a new model is applied, the Dyson-Schwinger (DS)
equation plays a predominantly important role in estimating the top mass.

The Bethe-Salpeter (BS) equation was derived originally by way of perturbation
by summing the series of repetitions of one loop. Schwinger instead has
derived this equation by the path integral formalism [3]. Such a
non-perturbative
approach based on the path integral has been largely developed since then,
and the procedure of the calculation is nowadays well formulated [4,5]. In this
approach the introduction of external fields (sources), denoted by
$\eta$ and $\bar\eta$ conventionally, is vitally important. The radiative
corrections appears through
 the sources $\eta$ and $\bar\eta$.

Meanwhile a new horizon is opened by Fujikawa for an aspect of the path
integral
[6]. The chiral
anomaly, whose signature was found for the first time by Fukuda,
Miyamota and Steinberger, and then was formulated by Adler, has been
dealt with by the path integral formalism in a concise manner.
The path integral measure is not invariant for the chiral transformation,
although
the Lagrangian is. As a consequence the Ward-Takahashi (WT) identity has
an anomalous term in comparison to what is expected by simple inspection
of the Lagrangian.

Fujikawa has developed further his method to deal with BRS symmetry
in QCD and the gravitational interaction where the non-linear transformations
take place [8].
Recently a more general kind of non-linear transformations, coherent
transformation so to speak,
was introduced and various new kinds of anomalous WT identities
with higher order products of fields have been deduced [9,10].
Higher the order of non-linearity, more involved the radiative corrections are.
Here again the use of external sources $\eta$ and $\bar\eta$
is indispensable to introduce desired radiative effects into the
identities. Therefore, if one carries out Schwinger's procedure in combination
with non-linear chiral
transformations, one ought to obtain equations which are more general than the
DS equations.

In the present article we wish to show that even the WT identity through
a simplest non-linear transformation includes an integrated form of the DS
equations
together with the field equations, and gives a more general form of the
relation
than the DS equation alone. To do this we adopt NJL, since the DS equation of
NJL
has been dealt with in a number of articles [2,11-16] such that we can compare
our result with theirs.

\tita{2. Procedure}

We start from the following non-linear chiral transformation [16]
$$\psi(x)\to\{(\exp \,\alpha(x)\,\gamma_5\,F(y,z))\,\psi(x)\}_+\eqno(1)$$
with
$$F(y,z)=\bar\psi(y)\,\gamma_5\,\psi(z)+\bar\psi(z)\,\gamma_5\,\psi(y).\eqno(2)$$

The suffix + indicates the chronological order.

Here a non-local form is chosen in order to bypass obstacles due to the
composite
fields. Then the  partition function relative to the NJL Lagrangian
$$L= i\bar\psi\spa{~}\psi+g\{(\bar\psi\psi)^2-(\bar\psi\gamma_5\psi)^2\}
\eqno(3)$$
changes by an infinitesimal amount as
$$\int \da\bar\psi\da\psi\,e^{i\int(L+\delta L+\alpha A) dx}\eqno(4)$$
in Fujikawa's notation. The quantity $\delta L=i\alpha F(y,z)\partial_\mu
(\bar\psi\gamma_5\gamma^\mu\psi)+O(\alpha^2)$ represents the variation of the
Lagrangian whereas the $A$ represents the variation
of the path integral measure and, in addition to the well-known chiral anomaly
$F(y,z)=\int \varphi_\lambda^+(x)\,\gamma_5\,\varphi_\lambda(x)\,d\lambda$,
contains anomalies of other kinds of tensors, and is
$$\eqalignno{
A&=\int d\lambda\{(\bar\psi(y)\gamma_5\psi(z))\,
 (\varphi_\lambda^+(x)\,\gamma_5\,\varphi_\lambda(x))\cr
&+\sum\limits_a {1\over N_a}\,(\bar\psi(y)J_a\psi(x))\,
( \varphi^+(x)\,J_a\,\varphi(z))\cr
&+\sum\limits_a {1\over N_a}\,(\bar\psi(x)J_a\psi(z))\,
( \varphi^+(y)\,J_a\,\varphi(x))\cr
&+(y\leftrightarrow z)&(5)\cr}$$
arranged in chronological order. The $J_a$ runs over all possible tensors,
1, $\gamma_\mu$,\hfill\break
 $\displaystyle{1\over2}$
$(\gamma_\mu\gamma_\nu-\gamma_\nu\gamma_\mu),$ $\gamma_5\gamma_\mu$
and $\gamma_5$. The second and third terms are due to the variation
of $F(y,z)$ and $1/N_a$ appear as a consequence of Fierz rearrangement
between $\delta F(y,z)$ and $\psi(x)$ and are equal to 1/4 in our case.
The $\varphi_\lambda$ are a complete orthogonal set of functionals
introduced by Fujikawa [6,8]. The equations for $\varphi_\lambda$ have never
been
given explicitly for the case of BRS or for the gravitation because they are
rather complicated.

In our case it is just
$$\,\gamma^0\Dbar\varphi_\lambda=\lambda\,\varphi_\lambda\eqno(6)$$
with
$$\Dbar=i\spa{~}+2g\{\bar\psi(x)\psi(x)-\gamma_5
(\bar\psi(x)\gamma_5\psi(x)\}.\eqno(7)$$
The simplest WT identity can be obtained by applying
$\delta/\delta\alpha$ on $Z$, but does not contain sufficient
radiative effects for our purpose. Therefore we introduce
sources and apply
$\delta/\delta\eta$
$\delta/\delta\bar\eta$
$\delta/\delta\alpha$ on the expression of the
generating functional $Z$,
$$Z=\int \da \bar\psi\da\psi\,e^{i\int(L+\bar\psi\eta+\bar\eta\psi)}\,dx.
\eqno(8)$$
transformed by means of (1).
The WT identity thus obtained does contain sufficient ingredients but this time
it is
full of terms irrelevant for our purposes.

To extract only the terms necessary for us
we construct a device. We first apply $\Dbar$ and $\bar\Dbar$ as
$$\Dbar^{cb}(\omega)\,\bar\Dbar^{ad}(u)
{\delta\over\delta\bar\eta^b(\omega)}\,
{\delta\over\delta\eta^a(u
)}\,
\left.{\delta\over\delta\alpha}\,Z\right\vert_{\alpha=0}=0\eqno(9)$$
where $\bar\Dbar=\gamma^0\Dbar^+\gamma^0$.
The $\Dbar(\omega)\,\delta/\delta\bar\eta(\omega)$ corresponds to
$\delta/\delta\bar\psi(\omega)$ in Schwinger's procedure
for deducing the DS equation, (see eq.(10.13) of [5]). If one
integrates over $u$, one gets a generalized form of the DS equation.

The $\bar\Dbar(u)\,\delta/\delta\eta(u)$, together with
 $\Dbar(\omega)\,\delta/\delta\bar\eta(\omega)$,
have been used also to derive the BS equation (see eq.10.24
of [5]).
If we had started from a transformation of a higher non-locality and
non-linearity differing by additional factor
$\{(\bar\psi(y')\psi(z'))^2-
(\bar\psi(y')\gamma_5\psi(z'))^2+(y'\rightarrow z')\}$,
the above identity (9) would have contained a generalized form
of the BS equation.

Now the eq.(9) itself is already too complicated and contains equations
other than what we seek. However there is an alternative way to obtain
a simple relation which the quark mass has to satisfy. When we apply
$\Dbar^{cb}(\omega)$ and $\bar\Dbar^{ad}(u)$ on
$(\delta/\delta\bar\eta^b(\omega)$
$\delta/\eta^a(u)$ $\delta/\alpha(x))Z$ in eq.(9), pairs
$\bar\psi(x)\;\psi(\omega)$ and
$\bar\psi(u)\;\psi(x)$
are reduced to $\delta$-functions. When integrated over $u$
and $\omega$, eq.(9) becomes very simplified.
All the irrelevant components drop out when one takes the trace due to
$\gamma_5$ matrices. We arrive at the following concise identity
$$\eqalignno{
g&<(\bar\psi(y)\gamma_5\psi(y)+(y\lrto z))
(\bar\psi(y)\gamma_5\psi(z)+(y\lrto z))\cr
&-(\bar\psi(y)\psi(y)+(y\lrto z))
(\bar\psi(y)\psi(z)+(y\lrto z))>\cr
&=<Tr\int\varphi_\lambda^+(y)\varphi_\lambda(z)d\lambda + (y\lrto z)>
&(10)\cr}$$
Here (in (5) also) we emphasize that the rule for the chronological order
is applied not only for $\psi$ and $\bar\psi$ but also for $\varphi_\lambda$
and $\bar\varphi_\lambda$ as if $\varphi_\lambda$ and $\bar\varphi_\lambda$
were fermionic field variables too. For example, in rhs, $<\varphi_\lambda
^+(y)\varphi_\lambda(z)>$ are $\varphi_\lambda^+(y)\varphi_\lambda(z)$ or $
-\varphi_\lambda(z)\varphi_\lambda^+(y)$ according to if $y_0>z_0$ or
$z_0>y_0$.

\tita{3. Terms from the path integral measure (the anomaly)}

Now we have to calculate rhs of (10) explicitly. The local one
$\int\varphi_\lambda^+(x)\varphi_\lambda(x)d\lambda$
appears in the dilatation anomaly. It diverges and has been
regularized by Fujikawa \ref(8). The
$\int\varphi_\lambda^+(y)\varphi_\lambda(z)d\lambda$
instead is regular, as will be seen below.

The simplest way to estimate the rhs of (10) seems to be to introduce a fifth
time $s$, called the proper time by Dirac, and to reexpress eq.(6) as
$$H\varphi_\lambda(x,s)=\lambda \varphi_\lambda(x,s)={id\over ds}
\varphi_\lambda(x,s)\eqno(11)$$
with the Hamiltonian $H=\gamma^0\Dbar$
which is the conjugate variable to $s$. Then  due to the relation
$\partial_{x_0}\theta(x_0-y_0)=\delta(x_0-y_0)$ the matrix with
chronologically ordered elements
$$B_a^b(y, s':x,s)=
\int\varphi_{a\lambda}^+(y,s')\varphi_\lambda^b(x,s) d\lambda\eqno(12)$$
obeys the equation
$$(\gamma^0\Dbar-i\partial_s)_b^c\,B_a^b(y, s':x,s)=
-i\delta_{ca}\delta^4(x-y)\,\delta(s-s')\eqno(13)$$
Its solution is
$$B(y, s':x,s)=
\int d\alpha\exp\left\lbrace i(H-i\partial_s)\alpha\right\rbrace
\delta^4(x-y)\,\delta(s-s')\eqno(14)$$
which, after some manipulation, can be brought into the following
form,
$${1\over(2\pi)^4}\int d^4k\,e^{-iky}e^{-iH(s'-s)}e^{ikx}\eqno(15)$$
(see Appendix A).

In the trace of the matrix $B$, only the terms with an even number
of $\gamma$ survive. Therefore one can save the labor by starting from the
second order equation
$$\eqalignno{
H_2\varphi_\eta(x,\tau)&=\eta\varphi_\eta(x,\tau)=
i\partial_\tau\varphi_\eta(x,\tau)\cr
&=i(\partial_s)^2\,\varphi_\lambda(x,s)&(16)\cr}$$
where in euclidian space
$$H_2=H^2= \Dbar^+ \Dbar =\left\lbrace\spa{~}-2g(S+\gamma_5P)\right\rbrace
\,\left\lbrace \spa{~}+2g(S-\gamma_5P)\right\rbrace\eqno(17)$$

and $\eta=\lambda^2$. In this space
the $S$ and $P$ stand for $\psi^+\gamma^4\psi$ and
$\psi^+\,\gamma^4\gamma_5\psi$
respectively. For the Wick's rotated $\gamma$ matrices we take the
convention of Fujikawa [6].
The $\tau$, the conjugate variable to $H_2$, is also
called the proper time, and has been used by Schwinger for a similar
purpose \ref(17). With this $H_2$, the trace of the matrix $B$ is
$$Tr\,B(x,y)={i\over (2\pi)^4}\,Tr
\int d^4k\,e^{-iky}e^{-iH_2(\tau-\tau')}e^{ikx}\eqno(18)$$
which, at $x=y$, is identical to Fujikawa's formula for the regularization
of the anomaly, if one replaces the Euclidean time $i(\tau-\tau')$ by
Fujikawa's $1/M^2$ \ref(18).

In order to integrate $Tr\,B(x,y)$, it is convenient to use a new variable
$k'=k+i(x-y)M^2/2$. Then $Tr\,B(x,y)$ becomes proportional to
$\exp\{-(x-y)^2M^2/4\}$ in Minkowsky space, and is small except at
$x-y\leq1/M$.
Furthermore it merges smoothly into the local value
$Tr\,B(x,x)$ as $y$ approaches $x$ (Appendix B).
For conciseness in the following, we limit ourselves to the local value.
Then the process of calculation is the same as for dilatation anomaly,
except that $M$ is large but is kept finite for the reason given later.

\tita{4. The $WT$ identity}

Developing $Tr\,B$ in terms of $M$ and $1/M$,
and retaining terms independent of $1/M$,
we obtain the following result for the identity (10) (see again
Appendix B),
$$\eqalignno{
g\lt S^2-P^2\gt=&{1\over 2\pi^2}\,\langle M^2g^2(S^2-P^2)\cr
&\quad+{1\over2}g^2(\partial_\mu S)(\partial^\mu S)
 -2g^4S^4\cr
&\quad-{1\over2}g^2(\partial_\mu P)(\partial^\mu P)
 -2g^4P^4\cr
&\quad+4g^4S^2P^2\rangle&(19)\cr}$$
where $S=\bar\psi(x)\psi(x)$ and $P=\bar\psi(x)\gamma_5\psi(x)$ as is
mentioned previously, but in the Minkowsky space.

As will be seen below, one can easily confirm that the above eq.(19)
is consistent with the formula for $S$ and $P$ derived in the past in
refs.\ref(11-13). We first mention that $M$ is equal to the Euclidean
cutoff momentum, denoted as $\Lambda$, as is shown on page~3678 of \ref(18).
Then, with the use of the gap equation, $2\pi^2g(S^2-P^2)-M^2g^2(S^2-P^2)$
is equal to $-m^2g^2(S^2-P^2)$, and the identity (19) is reduced to
$$\eqalignno{
0=&<m^2g^2(S^2-P^2)+{1\over2}g^2(\partial_\mu S)(\partial^\mu S)\cr
&\quad-2g^4S^4-{1\over2}g^2(\partial_\mu P)
(\partial^\mu P)-2g^4P^4+4g^4S^2P^2>&(20)\cr}$$

This is indeed identical to the sum of the two equations (12) and (13)
of ref.\ref(11), except for the total derivative terms which disappear
when integrated. Our idea is now found to be on the right road.

\tita{5. Relation to the DS equation and the BS amplitude}

Our remaining task is to clarify how the new equations 19 and 20 are related to
the DS equations and BS amplitude.
Our $S=(\bar\psi\psi)$ and $P=(\bar\psi\gamma_5\psi)$ contain not only
quark propagators, but also bound states (Higgs and $\pi$ in particular).
We separate these two kind of components. We first introduce the
Legendre transform of $G=\log\,Z$, denoted conventionally by $\Gamma$.
Replacing
$\psi$ and $\bar\psi$ in the lhs of eq.(10) by $-i\delta/\delta\bar\eta$ and
$+i\delta/\delta\eta$ acting on the Legendre transformed $Z$ respectively, we
obtain the following result,
$$\eqalignno{
\lt\bar \psi^a(y)(\bar\psi(y)\psi(y))\psi^b(z)\gt&=
S_{ca}(yy)S_{bc}(zy)-S_{cc}(yy)S_{ba}(zy)\cr
&+{}^4G_{ab}^1&(21)\cr}$$
where
$$iS_{ab}(xx')=\delta^2G/\delta\bar\eta^a(x)\delta\eta^b(x').\eqno(22)$$
The ${}^4G_{ab}^1$ is
$${}^4G_{ab}^1={\delta^4G\over\delta\eta^a(y)(\delta\eta^c(y)\delta\bar\eta^c(y)
)
\delta\bar\eta^b(z)}\eqno(23)$$
and is related to the vertex $^4\Gamma$ through the formula
$$\eqalignno{
&\delta^4G/\delta\bar\eta^f(t)\delta\eta^c(z)
\delta\bar\eta^b(y)\delta\eta^e(x)\cr
&=i\int S_{ff'}(tt')S_{bb'}(yy'){}^4\Gamma_{f'c'b'e'}(t'z'y'x')
S_{c'c}(z'z)S_{e'e}(x'x)
dx'dy'dz'dt'&(24)\cr}$$
with
$${}^4\Gamma_{f'c'b'e'}(t'z'y'x')=\delta{}^4\Gamma/\delta\bar\psi^{f'}(t')
\delta\psi^{c'}(z')
\delta\bar\psi^{b'}(y')
\delta\psi^{e'}(x')\eqno(25)$$
which is (C5) of the appendix C in the lowest order approximation.
 The ${}^4G^1$ represents the four leg processes
and still contains quark contributions in addition to those of the Higgs and
$\pi$.

We now adopt the spectral representation for the quark propagator,
$$S_{ab}(x,y)=\int{1\over\alpha(p)\pbar-\beta(p)}\,e^{ip(x-y)}\,dp
\eqno(26)$$
and assume that $\alpha(p)$ and $\beta(p)$ are independent of $p$, implying
that
the bubble approximation is introduced for the quark line. Choosing the sum of
momenta of two incoming quarks, $p_3+p_4=p_1+p_2=q$ as an independent
variable, we obtain (see Appendix D)
$$\eqalignno{
&S_{ac}S_{ca}+{}^4G^1=i{(2\pi)^{-8}\over (4\pi)^2} \,
g\alpha^4\log {\Lambda^2\over \mu^2}\,\cr
&4^2\int dr\,dq\left\lbrace1-3{r^2\over q^2-4m^2}+O\left({1\over q^2-4m^2}, r^2
\right)\right\rbrace&(27)\cr}$$
where $r=p_3-p_4$. The
$O\left({1\over q^2-4m^2}, r^2\right)$
indicates higher order terms in $1/(q^2-4m^2)$, includes further factors
such as $1/q^2$ etc., and has to be disregarded in the bubble approximation.

In the same way the first term of lhs of eq.(10) is
$$\eqalignno{
\lt\bar\psi^a(y)\bar\psi(y)\,\gamma_5\psi(y)(\gamma_5\psi(z)^b)\gt=&
S_{ca}^5(yy)S_{bc}^5(zy)\cr
&-S_{cc}^5(yy)S_{ba}^5(zy)\cr
&+ {}^4G_{ab}^5&(28)\cr}$$
with
$$^4G_{ab}^5=\delta^4G/\delta\eta^a(y)(\delta\eta^c(y)\delta\bar\eta^a(y))
(\gamma_5\delta\bar\eta(z))^b\eqno(29)$$
Corresponding to (27) we obtain
$$\eqalignno{-S_{ac}^5S_{ca}^5-{}^4G^5=&-{i(2\pi)^{-8}\over (4\pi)^2}\,
g\alpha^4\log {\Lambda^2\over \mu^2}\cr
&4^2\int dr\,dq\left\lbrace1-3{r^2\over q^2}+O\left({1\over q^2}, r^2
\right)\right\rbrace&(30)\cr}$$
The first terms in (27) and (30), which are too highly divergent to
be renormalized off, cancel each other in eqs.(10) and (20).
(see also ref.\ref(13)).

The second terms in (27) and (30), which we denote as $D_\phi$ and
$D_\pi$ respectively, are evidently the Higgs (mass of Higgs is twice of $m$)
and $\pi$ propagators,
and are exactly equal to the BS amplitudes (2,5) and (2,7) of \ref(14),
when one integrates over $\ell$ of $\Gamma$ ((A.4) of \ref(14))
from $\mu$ to $\Lambda$ (see also \ref(15)) (note that there appears a
difference by a factor $\log\Lambda^2/\mu^2$ from \ref(14) as the result
of the way of integration adopted here, as seen in (D1)).

The sum (27) plus (30) multiplied by $g^2$, is now simply  $D_\phi-D_\pi$,
$$g^2(S_{ac}S_{ca}- S_{ac}^5S_{ca}^5+ {}^4G^1-{}^4G^5)=-i
(D_\phi-D_\pi)\eqno(31)$$
The eqs.(21) and (22) are now expressed in terms of propagators as
$$g^2\lt S^2-P^2\gt=-g^2(S_{aa}S_{cc}- S_{aa}^5S_{cc}^5)-i(D_\phi-D_\pi)
\eqno(32)$$
which is consistent with the conventional expressions for $S$ and $P$
(for example see p.7 of \ref(15))
$$\eqalignno{
gS&=gv+\phi\cr
gP&=\pi&(33)\cr}$$
with $m=2g\,v$.

When expressed in terms of $v$, $\phi$ and $\pi$ by eq.(33),
our identity (20) represents
the vacuum condition for the effective potential to be at a minimum,
that is, zero. The result is consistent with the chiral scalar model from
the effective field approach \ref(11-15). Here,
$<(S^2-P^2)^2>$ in eq.(20)
is replaced by $(<S^2>-<P^2>)^2$ corresponding
to the bubble approximation.

Compared to this effective field approach, in Schwinger's path integral
formalism \ref(3,4 and 5) the result is always given in terms of
propagators only. Through the relation (32), the identity (20) presents
an equation for propagators of quark, $\phi$ and $\pi$. In our particular
case the identity becomes very simple as will be explained below.

While the identity (10) is a function of $y$ and $z$, the identity (20)
depends upon $x$ only, and consequently all the propagators including
$D_\phi$ and $D_\pi$ are constants. Owing to the gap equation
the quark propagator $S$ is $m/2g$
and $S^5$ vanishes. Then eq.(20) becomes simply
$$D_\phi={m^2\over8}={(gS)^2\over2}\quad\hbox{with }D_\pi=0,\eqno(34)$$
%$$\eqalignno{
%0=&\langle -m^2 (gv)^2+2(gv)^4\cr
%&+(-m^2-12(gv)^2)D_\phi-<{1\over2}\partial_\mu\phi\partial^\mu\phi
%+2 \phi^4>\cr
%&+(m^2-4(gvS)^2)D_\pi-<{1\over2}\partial_\mu\pi\partial^\mu\pi
%+2\pi^4>\cr
%&+4\pi^2\phi^2+4(D_\phi+ D_\pi)^2&(34)\cr}$$
stating again that the Higgs mass is twice of the quark mass in the bubble
approximation.

%$$\eqalignno{&<(\hbox{$\dal$}_y+4m^2+8\phi^2(x)-8\pi^2(x))
%i(\phi(x)\phi(y))>=\delta(x-y)\cr
%&<(\hbox{$\dal$}_y-8\pi^2(x)+8\phi^2(x))i(\pi(x)\pi(y))>=\delta(x-y)&(35)\cr}$$
%ignoring $D_\phi D_\pi$.

Thus our WT identity (10) is fully consistent with the result from the
effective
potential approach in the low energy domain \ref(12-14).

\vskip.9truecm\noi
{\bf Discussion}
\medskip
Throughout the above calculation we have assumed the cutoff momentum
$M$ to be large such that $O(1/M^2)$ is small. The higher dimensional
interaction terms which have been discussed vigorously in \ref(19)
and \ref(20) in particular, appear in the $O(1/M^2)$ in our treatment.
Evidently these terms have to be examined also, if a relatively small
cutoff momentum is chosen.
\vfill\eject

\vskip.9truecm\noi
{\bf Conclusion}
\medskip
We have seen that our identity leads to the same result as the effective
potential
approach \ref(11-16) in the low energy region. This situation is quite
different for events with large momentum transfer. In our treatment,
the Higgs is not inserted as an effective field $\phi$. Instead the Higgs's
propagator appears in the vertex function in a natural manner, and is expected
to behave quite differently from the result obtained
from the effective field $\phi$ for
very high energy events.

It has been known that the effective potential approach predicts results
precisely and rapidly for the low energy phenomena. On the contrary, we have
known
for some time that it is very dangerous to apply the effective theorems for
high energy processes \ref(21). In future experiments beyond the LEP energy
the effective potential approach will not always be reliable.

Our approach can handle such high energy problems since our identity is an
exact
relation. In this connection, it is interesting to compare, for example
the improved gap equation (2,74) of \ref(22) obtained from the refined
effective potential, to our identity. The improvement made in their eq.(2,74)
corresponds
to taking into account further higher terms $O(x,x,\,1/M^2)$ in our
(B4). However our $O(x,x,\,1/M^2)$ contains terms not appearing in the
effective
potential. It will be still more interesting to examine the auxiliary field
propagator of sect.~4 of ref.\ref(22) by inspection of our non-perturbative
formula and try to see how our identity predicts differently from their
refined effective potential at very high energy.

Also, in comparison with the orthodox Feynman-Dyson perturbation calculations,
our treatment can produce the results more quickly by saving an immense amount
of labor. Still,
further ingenuity will be needed to solve the identity as the energy
goes up higher.
\vfill\eject
\noi
{\bf Appendix A}
\bigskip
Expressing $\delta (s-s')$ as
$$\delta(s-s')={1\over 2\pi}\int_{-\infty}^\infty d\kappa\,e^{i\kappa(s-s')},
\eqno\rm(A1)$$
one can rewrite the solution (14) as
$$\eqalignno{
B(y, s':x,s)&={1\over 2\pi}\int d\kappa d\alpha\,e^{i(H-\kappa)\alpha}
\delta^4(x-y)\,e^{i\kappa(s-s')}\cr
&=\int d\alpha\delta(\alpha+s-s')\,e^{iH\alpha} \delta^4(x-y)\cr
&=\,e^{-iH(s-s')} \delta^4(x-y)\cr
&={1\over (2\pi)^4}\int d^4k\,e^{-iky}
e^{-iH(s-s')}\,e^{ikx}&\rm(A2)\cr}$$
which is the expression (15).
\vfill\eject
\noi
{\bf Appendix B}
\bigskip
First we rewrite $Tr\,B(x,y)$ (18)
$$
\int d^4k\,e^{-iky}e^{-\Dbar^+\Dbar/M^2} e^{ikx}
=\int d^4k\,e^{ik(x-y)}\,e^{-\Dbar^+(x,k)\Dbar(x,k)/M^2}
\eqno\rm(B1)$$
where $\Dbar(x,k)$ is defined as $\Dbar e^{ikx}f(\bar\psi,\psi)= e^{ikx}
\Dbar(x,k)f(\bar\psi,\psi)$, $f(\bar\psi,\psi)$ being a test functional.

\noi
There are two choices for the expression of the derivative operator
$\partial$ in $\Dbar(x,k)$ and $\Dbar^+(x,k)$. If one adopts
$\partial=\partial/\partial x$, we will get involved with a complicated
calculation. Alternatively we adopt the expression
$(\partial\psi/\partial x) \delta/\delta\psi+
(\partial \bar\psi/\partial x)\delta/\delta\bar\psi$
for $\partial$ in the present paper. The $\Dbar^+(x,k)\Dbar(x,k)$ commutes
with
$ik(x-y)$, and we get the result rapidly.
As a matter of fact, both choices should lead to the same result.

Then the (B1), that is,
$$e^{-\Dbar^+(x,k)\Dbar(x,k)/M^2+ik(x-y)}\eqno\rm(B2)$$
can be rewritten in terms of the translated momentum
$k'=k+i(x-y)M^2/2$ as
$$\eqalignno{
&\exp-\{(x-y)^2M^2/4\}\;\exp -\lbrace- k^{'2}+M^2(x-y)\partial+2ik'\partial\cr
&+\partial^2+
\spa{~} (2g)(S-\gamma_5P)-(2g)(S+\gamma_5P) \spa{~}-(2g)^2(S^2-P^2)\rbrace/M^2
&\rm(B3)\cr}$$
The presence of the terms $(x-y)\partial$ in the last expression implies
 the translational invariance of the result.

Because of the factor $\exp-\{(x-y)^2M^2/4\}$ the (B3) is small except when
$x\approx y$ and can be expressed as
$$e^{-(x-y)^2M^2/4} e^{-(x-y)\partial} \left(M^4+A(x,x)+O(x,x,1/M^2) +
O(x-y)\right),\eqno\rm(B4)$$
after the integration over $k'$ from 0 to $M$.

Among the terms of the power series in $1/M^2$, $A(x,x)$ represents terms
independent of $M^2$, and the $O(x,x,1/M^2)$ indicate the rest.

The $O(x-y)$ is a power series in $(x-y)$ and arises as a result of
the factorization of $\exp-\{(x-y)\partial\}$. The final result for $A(x,x)$
is in the Minkowski space
$$\eqalignno{
A(x,x)=&{-1\over\pi^2} \lbrace M^2g^2(S^2-P^2)\cr
&+{1\over2} g^2 (\partial_\mu S)(\partial^\mu S)-2g^4S^4\cr
&-{1\over2} g^2 (\partial_\mu P)(\partial^\mu P)-2g^4P^4\cr
&+4g^4 S^2 P^2\rbrack\rbrace&\rm(B5)\cr}$$
with $S=\bar\psi(x)\psi(x)$ and $P=\bar\psi(x)\gamma_5\psi(x)$, up to total
derivatives.

The first term proportional to $M^4$ in (B4) is independent of the interaction,
and has
to be normalized off. The rest of (B4) approaches smoothly $A(x,x)$ as $y$
approaches $x$.
\vfill\eject

\noi
{\bf Appendix C}
\bigskip
There are several ways to obtain $^4G^1$ in terms of the Legendre
transform $\Gamma$. We have started from the relation
$$\eqalignno{
{\delta\bar\psi^{b'}(y')\over\delta\bar\psi^{a'}(x')}
&=\delta_{a'b'}\delta(x'-y')=\int
{\delta\bar\eta^{e'}(x'')\over\delta\bar\psi^{a'}(x')}
{\delta\bar\psi^{b'}(y')\over\delta\bar\eta^{e'}(x'')}dx''\cr
&=\int{\delta^2\Gamma\over
\delta\bar\psi^{a'}(x')\delta\psi^{e'}(x'')}
i{\delta^2 G\over
\delta\bar\eta^{e'}(x'')\delta\eta^{b'}(y')}dx''&\rm(C1)\cr}$$

\noi
Applying
$\delta/\delta\eta^{c'}(z')$ and $\delta/\delta\bar\eta^{f'}(t')$,
we arrive at the following relation
$$\eqalignno{
&-\int S_{c''c'}(z''z') S_{f'f''}(t't''){\delta^4\Gamma\over
\delta\bar\psi^{f''}(t'')\delta\psi^{c''}(z'')
\delta\bar\psi^{a'}(x')\delta\psi^{e'}(x'')}
i S_{e'b'}(x''y')dx''dz''dt''\cr
&=\int\Gamma_{a'e'}^2(x'x'')
{\delta^4 G\over
\delta\bar\eta^{f'}(t')\delta\eta^{c'}(z')
\delta\bar\eta^{e'}(x'')\delta\eta^{b'}(y')}dx''
&\rm(C2)\cr}$$
from which we obtain $^4G^1$ as
$$\eqalignno{
&{\delta^4 G\over
\delta\bar\eta^f(t)\delta\eta^c(z)
\delta\bar\eta^b(y)\delta\eta^{e}(x)}
=i\int S_{ff'}(tt') S_{bb'}(yy')\cr
&{\delta^4\Gamma\over
\delta\bar\psi^{f'}(t')\delta\psi^{c'}(z')
\delta\bar\psi^{b'}(y')\delta\psi^{e'}(x')}
 S_{c'c}(z'z)S_{e'e}(x'x) dx' dy'dz'dt'
&\rm(C3)\cr}$$

\noi
where
$$\Gamma_{a'e'}^2(x'x)=\delta^2\Gamma/
\delta\bar\psi^{a'}(x')\delta\psi^{e'}(x)
\eqno\rm(C4)$$

\noi
is the inverse of the propagator.
In the lowest order, the vertex function is simply
$$\eqalignno{
&^4\Gamma^1_{f'c'b'e'}(t'z'y'x')\cr
&=\delta ^4\Gamma/
\delta\bar\psi^{f'}(t')\delta\psi^{c'}(z')
\delta\bar\psi^{b'}(y')\delta\psi^{e'}(x')\cr
&=2g(\delta_{c'f'}\delta_{e'b'}-\delta_{c'b'}\delta_{e'f'})
\delta(x'-t')\delta(x'-z')\delta(x'-y')
&\rm(C5)\cr}$$

\noi
$^4G^5$ can be obtained in a similar manner.
\vfill\eject

\noi
{\bf Appendix D}
\bigskip
We choose the sum of momenta of two incoming particles, say $p_1+p_2=q$,
as an independent variable, because $q$ represents the momentum of the
bound states, that is, the Higgs and $\pi$ (we ignore vector and
pseudo vector bound states). Let us denote the momenta of four
$S_{ab}$ in $^4G^1$ as $p_1$, $p_2$, $p_3$ and $p_4$ successively.
Expressing $p_1$ as $p_1=q-p_2$, and integrating over $p_2$,
from $\mu$ to $\Lambda$ we obtain
$$\eqalignno{&^4G^1=i4(2\pi)^{-8}\int\left\lbrace{1\over8}+
{\alpha^2g\log{\Lambda^2\over \mu^2}\over (4\pi)^2}
(4m^2-q^2)\right\rbrace\cr
&\left( Tr{1\over \pbar_3-m}\;{1\over\pbar_4-m}\right)\;dp_3\,dp_4\,dq
&\rm(D1)\cr}$$

\noi
The first term of $^4G^1$ is equal to
$S_{ac} S_{ca}$. Here the gap equation has already been used
for this term in order to eliminate $g$. ( We did not employ the integration
used by BHL who take into account the scaling aspect.)

The last factor of $^4G^1$, namely the product of the last two
$S_{ab}$, can be expanded in terms of $q^2-4m^2$. The result for the second
term of $^4G^1$ is
$$\eqalignno{
&\int(4m^2-q^2)Tr{1\over\pbar_3-m}\;{1\over\pbar_4-m}\;dp_3\,dp_4\cr
&=\int dr 4^2\left.\left(1-{r^2\over q^2-4m^2}+{8m^2\over
(q^2-4m^2)^2}\right)\right/\cr
&\left(1+{2(qr)+r^2\over q^2-4m^2}\right)
\left(1+{-2(qr)+m^2\over q^2-4m^2}\right)\cr
&=\int dr 4^2\left\lbrace1-{3r^2\over q^2-4m^2}+O\left({1\over
q^2-4m^2},\,r^2\right)\right\rbrace&\rm(D2)\cr}$$

\noi
In the same way $^4G^5$ is
$$\eqalignno{&^4G^5=i4(2\pi)^{-8}\int\left\lbrace{1\over8}-
{\alpha^2g\log{\Lambda^2\over \mu^2}\over (4\pi)^2}
q^2\right\rbrace\cr
&\left( Tr\,\gamma_5{1\over(\pbar_3-m)}\,\gamma_5\,
{1\over\pbar_4-m}\right)\;dp_3\,dp_4\,dq
&\rm(D3)\cr}$$

\noi
The first term is equal to
$S_{ac}^5 S_{ca}^5$.
For the second term we have
$$\eqalignno{
&\int q^2 Tr\left(\gamma_5{1\over(\pbar_3-m)}\,\gamma_5\,
{1\over\pbar_4-m}\right)\;dp_3\,dp_4\cr
&=-\int dr 4^2\left\lbrace1-{3r^2\over q^2}+O\left({1\over
q^2},\,r^2\right)\right\rbrace&\rm(D4)\cr}$$

\vskip 2truecm
\noindent
REFERENCES
\bigskip

\aut Y. Nambu and G. Jona-Lasinio*
\editeur Phys. Rev.* \gras 122* (1961)~345.
\smallskip

\aut Y. Nambu* : \gui BCS Mechanism, Quasi-Supersymmetry, and Fermion Mass
Matrix"*, talk presented at the Kasimierz Conference, EFI 88-39
(July 1988); see also International Workshop reports of 1988 and 1989,
Nagoya, Japan :
{\it Ed. Bando, Muta and Yamawaki}\hfill\break
Also there are a number of pioneer works on the top condensate
by Nagoya group, K.~Yamawaki in particular, around 1989-1990. The most
of early ones are published in the Modern Physics Letters, the Modern
Physics and reports of international workshops held at Nagoya from 1988
to 1990 :
{\it (Ed. Yamawaki et all, World Scientific).}
\smallskip

\aut J. Schwinger*
\editeur Proc. National Academy of Sciences* \gras 37* (1951)~452.
\smallskip

\aut* As clear illustrations from the perturbative viewpoint, one may
consult {\smc K.~Nishijima}: Fields and Particles, New York
\editeur Benjamin, Inc.* 1969.
\suitaut T. Muta* : Foundation of Quantum chromodynamics,
Singapore, World Scientific, 1987.\hfill\break
For the non-perturbative aspect, we cite the latest book by
{\it J. Zinn-Justin}: Quantum Field Theory and Critical Phenomena,
New York, Oxford University Press 1990.
\smallskip

\aut* For the derivation of DS and BS equations see for example chapter~10
of C.~Itzykson and J.B.~Zuber: Quantum Field Theory, Singapore,
Mcgraw-Hill, 1986.
\smallskip

\aut K. Fujikawa*
\editeur Phys. Rev.* \gras D~21* (1980)~2848.
\smallskip

\aut H. Fukuda and Y. Miyamoto*
\editeur Proc. Theor. Phys.* \gras 4* (1949)~235,347;
\suitaut S. Adler*
\editeur Phys. Rev.* \gras 177* (1969)~2426.
\smallskip

\aut* There exist a number of articles by Fujikawa. As a summary talk for
early works see p.106 of Quantum Gravity and Cosmology
\editeur ed. H. Sato and T. Inami,* Singapore, World Scientific, 1986.
\smallskip

\aut J.L. Jacquot*
\editeur Phys. Lett.* \gras B212* (1988)~231,
\suitaut J.L. Jacquot and M. Umezawa*
\editeur Z. Phys.* \gras C45* (1990)~381.
\smallskip

\aut J.L. Jacquot and J. Richert*
\editeur Z. Phys.* \gras C56* (1992)~201.
\smallskip

\aut T. Eguchi and H. Sugawara*
\editeur Phys. Rev.* \gras D 10* (1974)~4257.
\smallskip

\aut K. Kikkawa*
\editeur Prog. Theor. Phys.* \gras 56* (1976)~947.
\smallskip

\aut T. Eguchi*
\editeur Phys. Rev.* \gras D 14* (1976)~2755.
\smallskip
\hfuzz=3pt
\aut W.A. Barden, C.T. Hill and M. Lintner*
\editeur Phys. Rev.* \gras D 41* (1990)~1647.
\smallskip

\aut C.T. Hill:* Top Quark Condensates,
unpublished, Fermi-Pub-92/19-T, Jan.15, 1992.
\smallskip

\aut* The point of view on the symmetry breaking resulting from this
transformation has been presented by us p.359 of the Bregenz Symposium
\gui Symmetry in Science VI" 1992, ed. B. Gruber, Plenum Press, New York 1993.
Here only the mass generation of quarks is treated in the ladder approximation.
All terms which do not contribute to the quark mass are ignored. The
Bethe-Salpeter amplitudes for bound states $\phi$ and $\pi$ (our $D_\phi$
and $D_\pi$) are neglected too.
\smallskip

\aut J. Schwinger*
\editeur Phys. Rev.* \gras 82* (1951)~664.
\smallskip

\aut M.~Umezawa*
\editeur Phys. Rev.* \gras D 39* (1989)~3672.
\smallskip

\aut M.~Suzuki*
\editeur Mod. Phys. Lett.* \gras A5* (1990)~1205;
see also {\smc W.A. Bardeen} \gui Electroweak Symmetry breaking:
Top Quark Condensates", the 5th Nishinomiya Yukawa Memorial
Symposium, Oct.~25 (1990).
\smallskip

\aut A. Hasenfratz, P. Hasenfratz, K. Jansen, J. Kuti, Y. Shen:*
\gui The Equivalence of the Top Quark Condensate and the Elementary
Higgs Field", USCD/PTH 91-06 (1991).
\smallskip

\aut * As one of the later examples, we cite the $Z\to 2\gamma$ decay. See
{\smc T. Hatzuda, M. Umezawa}:
{\it Phys. Lett.} \gras B254* (1991)~493. All other references are
cited therein.
\suitaut T. Hatzuda, M. Umezawa*:
The Decay $Z\to\sigma\gamma$, CRN Report (unpublished 1990);
The $Z\to\sigma\mu^+\mu^-$ as a test of the usage of PCDC in high energy
phenomena,  CRN Report (unpublished 1991); The Decay of $Z$ to a
Dialton and a $\mu$ pair: PCDC and the Decay Form Factor,
 CRN Report (unpublished 1992).
\smallskip

\aut K. Kondo, M. Tanabashi, K. Yamawaki:*
\gui Renormalization in the Gauged Nambu-Jona-Lasinio Model",
unpublished, KEK-TH-344, DPNU-92-23.
\smallskip

\bye